\documentclass[aps,showpacs,preprintnumbers,amsmath,amssymb]{revtex4}

\usepackage{dcolumn}
\usepackage{graphicx}
\usepackage[dvips]{epsfig}
\usepackage{amssymb}
\usepackage{color}
\usepackage{enumerate}
\usepackage{subfigure}
\usepackage{multirow}
\usepackage{diagbox}


\begin{document}
\baselineskip=0.8 cm
\title{\bf Shadow of a circular disformal Kerr black hole beyond GR}

\author{Fen Long$^{1}\footnote{lf@usc.edu.cn}$,
Songbai Chen$^{2,3}$\footnote{Corresponding author: csb3752@hunnu.edu.cn},
Jiliang Jing$^{2,3}$ \footnote{jljing@hunnu.edu.cn}}
\affiliation{ $ ^1$ School of Mathematics and Physics, University of South China,
Hengyang, 421001, People's Republic of China
\\$ ^2$ Department of Physics, Institute of Interdisciplinary Studies, \
\\ Key Laboratory of Low Dimensional Quantum Structures and Quantum Control of Ministry of Education,
\\ Synergetic Innovation Center for Quantum Effects and Applications, Hunan
Normal University,  Changsha, Hunan 410081, People's Republic of China
\\$ ^3$ Center for Gravitation and Cosmology, College of Physical Science and Technology, Yangzhou University, Yangzhou 225009, People's Republic of China}

\begin{abstract}
\baselineskip=0.6 cm
\begin{center}
{\bf Abstract}
\end{center}

We have studied the shadows of a circular disformal Kerr black hole with a deformation parameter, which represents a rotating solution within a generalized scalar-tensor framework of Horndeski gravity that characterized by second-order field equations. Our result show that for the non-rotating case, the shadow remains perfectly circular, with its radius independent of the deformation parameter. For the rotating case, the size of the shadow decreases as the deformation parameter decreases, and the shape of the shadow gradually becomes more flattened.
It is worth noting that while the shadow of a rotating black hole remains north-south symmetric for equatorial observers, this symmetry is broken once the observer moves away from the equatorial plane. 
For an observer in the northern hemisphere, the geometric center of the shadow shifts northward when $D_0<0$ and southward when $D_0>0$, while the opposite behavior occurs for observers in the southern hemisphere.
These features in the black hole shadow originating from the scalar field could help us to understand the circular disformal Kerr black hole and generalized scalar-tensor framework of Horndeski gravity.

\end{abstract}

\maketitle
\newpage
\section{Introduction}

The first images of the supermassive black holes in M87\cite{EHT1} and the Milky Way Galaxy\cite{EHT2}, obtained by the Event Horizon Telescope (EHT), have opened an unprecedented observational window into the strong-gravity regime. These observations provide direct evidence for the existence of compact objects consistent with the predictions of General Relativity(GR). Although GR has successfully passed current observational tests, an important question remains as to whether future high-resolution astronomical observations will continue to support its predictions. This consideration has motivated extensive research on black hole solutions and their observational signatures within various modified gravity frameworks.

It is well known that generalized scalar-tensor theories are a kind of alternative theories of gravity\cite{d36,d37,d38,d39,d40,d41,d42,dhosts10}, in which scalar fields participate in the gravitational interaction together with the metric tensor. The new black hole solutions generated from generalized scalar-tensor theories can be classified as the stealth solutions and the non-stealth solutions. For the stealth solutions, non-trivial scalar fields without backreacting on the metric then the metric of spacetime is the same as those in general relativity. However, for the non-stealth solutions, the parameters of scalar field appear in the metric, which lead to that the metric for the solution deviate from those in Einstein's theory of gravity. From an observational perspective, non-stealth black hole solutions are of particular interest, as scalar fields may leave detectable signatures in black hole shadows, gravitational-wave signals, and other astrophysical phenomena. The non-stealth solutions can be generated by applying disformal transformations to known ``seed'' solutions. For example, starting from a known ``seed'' solution in Degenerate Higher-Order Scalar-Tensor(DHOST) Ia theory, one may construct a new solution in another specific DHOST Ia theory by performing a disformal transformation of the metric\cite{d53,d54}. The resulting disformal Kerr spacetime represents a fully backreacting rotating solution and features a scalar field with both radial dependence and linear time evolution\cite{d51,d}. However, the disformal Kerr black hole in quadratic DHOST theories does not satisfy the circularity condition and difficult to clearly identify key geometrical features such as the Killing horizon and the ergoregion. More recently, a new fully backreacting rotating black holes with scalar hair has been constructed\cite{d}, distinguished by the fact that circularity is preserved. This solution is obtained from a Kerr stealth black hole in Horndeski theory which is restricted to purely quadratic sector. Moreover, key geometrical structures in the new disformal solution, such as the Killing horizon and the ergoregion, can be identified in the same way as in the Kerr black hole. Therefore, this new exact solution provides an improved version of the disformal Kerr black hole with excellent geometric properties, and the newly introduced a disformal parameter will also bring some new physical features to the spacetime that are different from Kerr case.

Images of objects encode valuable information about the spacetime\cite{sw,swo,astro,chaotic,binary,sha18,my,BI,swo7,swo8,swo9,swo10,swo11,swo12,S11,S12,W7,W8,W9,W10,W11,W12,W13,kns,kds,whk1,whk2}, making them a powerful tool for testing various modified theories of gravity\cite{W19,W20,W21,W22,W23,W24,S7,S8,S9,S10,BDKS,Z1,Z2,Z3}. Moreover, black hole shadows in the images have also been treated as a potential tool to study the possibility of constraining black hole parameters and extra dimension size\cite{extr1,extr2}, and to probe some fundamental physics issues including dark matter\cite{tomoch,dark1,dark2,dark3,dark4} and the equivalence principle\cite{epb}. In this paper, we will study the shadows of the circular disformal Kerr black hole and find new features exist in the shadow for this disformed black hole.

The paper is organized as follows. In Sec. II, we briefly review the circular disformal Kerr black hole beyond GR and and study equation of motion for the photons in this spacetime. In Sec. III, we present numerically the shadow of the circular disformal Kerr black hole and probe the effects of the deformation parameter arising from scalar field on the shadow.

\section{The circular disformal Kerr black hole beyond GR}

We now briefly introduce the circular disformal Kerr black hole, obtained from a Kerr stealth configuration through a disformal transformation. The stealth solution arises in a class of Horndeski theories restricted to their purely quadratic sector \cite{d,d62}, and its dynamics are governed by the action
\begin{equation}
S=\int \sqrt{-g} \mathrm{~d}^4 x\left\{G_4(X) R+G_{4 X}\left[(\square \varphi)^2-\left(\nabla_\mu \nabla_\nu \varphi\right)^2\right]\right\},
\end{equation}
where $R$ denotes the Ricci scalar, $\varphi$ is the scalar field with kinetic term $X=-\frac{1}{2} g^{\mu \nu} \partial_\mu \varphi \partial_\nu \varphi$ ,and the function $G_4(X)$ is required to satisfy the conditions $G_{4X}(X_0)=0$ and $G_{4XX}(X_0)=0$, with $G_{4X}(X)=\partial G_4 / \partial X$. Hence, for any theory of the form
\begin{equation}
G_4(X)=\kappa+\sum_{i \geq 2} \beta_i\left(X-X_0\right)^i,
\end{equation}
there exists a General Relativity vacuum solution supporting a non-trivial scalar field profile determined by the constraint $(X = X_0)$. The scalar field corresponding to the Kerr stealth solution is given by
\begin{equation}
\varphi(r, \theta)=\sqrt{-2 X_0}\bigg[a \sin \theta-\sqrt{\Delta}-m \ln (r-m+\sqrt{\Delta})\bigg],
\end{equation}
and its gradient is expressed as
\begin{equation}
\partial_r \varphi=-r \sqrt{\frac{-2 X_0}{\Delta}}, \quad \partial_\theta \varphi=\sqrt{-2 X_0} a \cos \theta . \label{varphi}
\end{equation}
One can obtain a new solution from a ``seed'' known solution by performing a disformal transformation of the metric. In general, the  disformal transformation of the metric can be expressed as \cite{dhosts10}
\begin{equation}
g_{\mu \nu} = C(\varphi, X) \tilde{g}_{\mu \nu}+D(\varphi, X) \partial_\mu \varphi \partial_\nu \varphi,\label{transformation}
\end{equation}
where $g_{\mu \nu}$ is the ``disformed" metric and $\tilde{g}_{\mu \nu}$ is the original ``seed" one. The functions $ C(\varphi, X)$ and $D(\varphi, X)$ denote the conformal and disformal factors, respectively, which for simplicity are taken to be constants, $ C(\varphi, X)=C_0$ and $D(\varphi, X)=D_0$. Combining Eqs.(\ref{varphi}) and (\ref{transformation}), the Kerr solution after the disformal transformation can be written in the following form\cite{d}:

\begin{equation}
\begin{aligned}
\mathrm{d} s^2 & =C_0\left[-\left(1-\frac{2 M r}{\Sigma}\right) \mathrm{d} t^2-\frac{4 a M r \sin ^2 \theta}{\Sigma} \mathrm{~d}t \mathrm{~d}\psi+\left(r^2+a^2+\frac{2 a^2 M r \sin ^2 \theta}{\Sigma}\right) \sin ^2 \theta \mathrm{~d} \psi^2\right] \\
& +\Sigma\left[\left(\frac{C_0}{\Delta}-\frac{2 D_0 X_0 r^2}{\Delta \Sigma}\right) \mathrm{d} r^2+\left(C_0-\frac{2 D_0 X_0 a^2 \cos ^2 \theta}{\Sigma}\right) \mathrm{d} \theta^2+\frac{4 D_0 X_0}{\sqrt{\Delta} \Sigma} a r \cos \theta \mathrm{~d} r \mathrm{~d} \theta\right],
\end{aligned} \label{metric}
\end{equation}
with
\begin{equation}
\begin{aligned}
\Sigma=r^2+a^2 \cos ^2 \theta, \quad  \Delta & =r^2+a^2-2 M r,
\end{aligned}
\end{equation}
where $M$ and $a$ represent the mass and the rotation parameter, respectively.
Unlike the disformal Kerr black holes in quadratic DHOST theories \cite{d53,d54}, this solution preserves circular symmetry \cite{d}, thereby ensuring that the spacetime is free of closed timelike curves.
This solution is generated from the Kerr stealth black hole within Horndeski gravity, whose associated scalar field continues to respect stationarity and axisymmetry\cite{d62}.
By applying a disformal transformation compatible with these symmetries, extra convective effects are avoided, and the circular structure of the spacetime is preserved\cite{d}.
This circular disformal Kerr metric has the same geometric properties as the Kerr black hole, such as the event horizon, ergospheres, and ring singularity.
The event horizon $r_{+}$ of the circular disformal Kerr metric are given by the root of the following equation
\begin{eqnarray}
\Delta = r^2+a^2-2 M r= 0.
\end{eqnarray}
However, the metric exhibits a nonzero $g_{r\theta}$ component, indicating that the scalar field now affects the spacetime geometry and that the circular disformal Kerr metric is no longer a stealth solution. Moreover, due to the presence of the $g_{r\theta}$ term, its asymptotic behavior is no longer entirely identical to that of the Kerr metric.

We now examine the motion of photons in the circular disformal Kerr spacetime (\ref{metric}). In a curved spacetime, the Hamiltonian for photons propagating along null geodesics can be expressed as
\begin{eqnarray}
 H(x,p)=\frac{1}{2}g^{\mu \nu}(x)p_{\mu}p_{\nu}=0.\label{hamiltonian}
\end{eqnarray}
There are two conserved quantities: the energy $E$ and the angular momentum $L_z$ with the following expressions
\begin{eqnarray}
E=-p_{t}=-g_{tt}\dot{t}-g_{t\psi}\dot{\psi}, \quad L_{z}=p_{\psi}=g_{t\psi}\dot{t}+g_{\psi\psi}\dot{\psi}.\label{conserved quantities}
\end{eqnarray}
From these conserved quantities we obtain the photon equations of motion along null geodesics
\begin{eqnarray}
\dot{t}\quad=&&\frac{g_{\psi\psi}E+g_{t\psi}L_z}{g_{t\psi}^2-g_{tt}g_{\psi\psi}},\label{u1}\\
\dot{\psi}\quad=&&\frac{g_{t\psi}E+g_{tt}L_z}{g_{tt}g_{\psi\psi}-g_{t\psi}^2},\label{u4}\\
\ddot{r}\quad=&&\frac{1}{2(g_{rr}g_{\theta\theta}-g_{r\theta}^2)}\bigg\{g_{\theta\theta}\bigg[g_{tt,r}\dot{t}^2-g_{rr,r}\dot{r}^2-2g_{rr,\theta}\dot{r}\dot{\theta}
+(g_{\theta\theta,r}-2g_{r\theta,\theta})\dot{\theta}^2+g_{\psi\psi,r}\dot{\psi}^2+2g_{r\psi,r}\dot{t}\dot{\psi}\bigg]\nonumber\\
&&-g_{r\theta}\bigg[g_{tt,\theta}\dot{t}^2
+(g_{rr,\theta}-2g_{r\theta,r})\dot{r}^2-2g_{\theta\theta,r}\dot{r}\dot{\theta}-g_{\theta\theta,\theta}\dot{\theta}^2+g_{\psi\psi,\theta}\dot{\psi}^2
+2g_{t\psi,\theta}\dot{t}\dot{\psi}\bigg]\bigg\},\label{uu2}\\
\ddot{\theta}\quad=&&\frac{1}{2(g_{r\theta}^2-g_{rr}g_{\theta\theta})}\bigg\{g_{r\theta}\bigg[g_{tt,r}\dot{t}^2-g_{rr,r}\dot{r}^2-2g_{rr,\theta}\dot{r}\dot{\theta}
+(g_{\theta\theta,r}-2g_{r\theta,\theta})\dot{\theta}^2+g_{\psi\psi,r}\dot{\psi}^2+2g_{r\psi,r}\dot{t}\dot{\psi}\bigg]\nonumber\\
&&-g_{rr}\bigg[g_{tt,\theta}\dot{t}^2
+(g_{rr,\theta}-2g_{r\theta,r})\dot{r}^2-2g_{\theta\theta,r}\dot{r}\dot{\theta}-g_{\theta\theta,\theta}\dot{\theta}^2+g_{\psi\psi,\theta}\dot{\psi}^2
+2g_{t\psi,\theta}\dot{t}\dot{\psi}\bigg]\bigg\}.\label{uu3}
\end{eqnarray}
From Eqs.(\ref{uu2}) and (\ref{uu3}), it can be seen that $\ddot{r}$ and $\ddot{\theta}$ both contain the $g_{r\theta}$ term, which means that the motion of photon has some behaviors differed from the case of Kerr black hole. Thus, it is expected that the shadow of the circular disformal Kerr black hole(\ref{metric}) should possess some new properties which do not belong to Kerr one.

\section{The images of the circular disformal Kerr black hole beyond GR}
In the following, we investigate image formation in the circular disformal Kerr black hole and examine the influence of the deformation parameter $D_0$ on the images.
We employ the ``backward ray-tracing" method\cite{sw,swo,astro,chaotic,binary,sha18,my,BI,swo7,swo8,swo9,swo10} to numerically simulate the shadow of the circular disformal Kerr black hole. In this approach, light rays are traced backward from the observer by numerically integrating the null geodesic equations(\ref{u1})-(\ref{uu3}), which determines the position of each pixel in the final image. The observer's basis $\{e_{\hat{t}},e_{\hat{r}},e_{\hat{\theta}},e_{\hat{\psi}}\}$ can be expressed in terms of the coordinate basis$\{\partial_t,\partial_r,\partial_{\theta},\partial_{\psi} \}$.
\begin{eqnarray}
\label{zbbh}
e_{\hat{\mu}}=e^{\nu}_{\hat{\mu}} \partial_{\nu},
\end{eqnarray}
where the matrix $e^{\nu}_{\hat{\mu}}$ satisfies $g_{\mu\nu}e^{\mu}_{\hat{\alpha}}e^{\nu}_{\hat{\beta}}
=\eta_{\hat{\alpha}\hat{\beta}}$, and $\eta_{\hat{\alpha}\hat{\beta}}$ denotes the usual Minkowski metric. For the circular disformal Kerr black hole (\ref{metric}), it is convenient to choose a decomposition \cite{sw,swo,astro,chaotic,binary,sha18,my,BI,swo7,swo8,swo9,swo10,swo11,swo12}
\begin{eqnarray}
\label{zbbh1}
e^{\nu}_{\hat{\mu}}=\left(\begin{array}{cccc}
\zeta&0&0&\gamma\\
0&\eta&\varepsilon&0\\
0&0&A^{\theta}&0\\
0&0&0&A^{\psi}
\end{array}\right),
\end{eqnarray}
where $\zeta$, $\gamma$, $\eta$, $\varepsilon$, $A^{\theta}$,and $A^{\psi}$ are real coefficients.
From the Minkowski normalization condition
\begin{eqnarray}
e_{\hat{\mu}}e^{\hat{\nu}}=\delta_{\hat{\mu}}^{\hat{\nu}},
\end{eqnarray}
it follows that
\begin{eqnarray}
\label{xs}
&&\zeta=\sqrt{\frac{g_{\psi\psi}}{g_{t\psi}^{2}-g_{tt}g_{\psi\psi}}},\;\;\;\;\;
\gamma=-\frac{g_{t\psi}}{g_{\psi\psi}}\sqrt{\frac{g_{\psi\psi}}{g_{t\psi}^{2}-g_{tt}g_{\psi\psi}}},\;\;\;\;\;
\eta=\sqrt{\frac{g_{\theta\theta}}{g_{rr}g_{\theta\theta}-g_{r\theta}^2}},\nonumber\\
&&\varepsilon=-\frac{g_{r\theta}}{\sqrt{g_{\theta\theta}(g_{rr}g_{\theta\theta}-g_{r\theta}^2})},\;\;\;\;\;
A^{\theta}=\frac{1}{\sqrt{g_{\theta\theta}}},\;\;\;\;\;\;\;
A^{\psi}=\frac{1}{\sqrt{g_{\psi\psi}}}.
\end{eqnarray}
According to Eq.(\ref{zbbh}), the locally measured four-momentum $p^{\hat{\mu}}$ of a photon takes the form
\begin{eqnarray}
\label{dl}
p^{\hat{t}}=-p_{\hat{t}}=-e^{\nu}_{\hat{t}} p_{\nu},\;\;\;\;\;\;\;\;\;
\;\;\;\;\;\;\;\;\;\;\;p^{\hat{i}}=p_{\hat{i}}=e^{\nu}_{\hat{i}} p_{\nu}.
\end{eqnarray}
By making use of Eq.(\ref{xs}), the locally measured four-momentum $p^{\hat{\mu}}$ in the circular disformal Kerr black hole can be derived as follows
\begin{eqnarray}
\label{kmbh}
p^{\hat{t}}&=&\zeta E-\gamma L,\;\;\;\;\;\;\;\;\;\;\;\;\;\;\;\;\;\;\;\;
p^{\hat{r}}=\eta p_{r}+\varepsilon p_{\theta},\nonumber\\
p^{\hat{\theta}}&=&\frac{1}{\sqrt{g_{\theta\theta}}}p_{\theta},
\;\;\;\;\;\;\;\;\;\;\;\;\;\;\;\;\;\;\;\;\;\;
p^{\hat{\psi}}=\frac{1}{\sqrt{g_{\psi\psi}}}L,
\end{eqnarray}
Hence, the celestial coordinates corresponding to a given light ray in the spacetime (\ref{metric}) are written as
\begin{eqnarray}
\label{xd1}
\alpha&=&-r_{obs}\frac{p^{\hat{\psi}}}{p^{\hat{r}}}
=-r_{obs}\frac{1}{\sqrt{g_{\psi\psi}}}\frac{g_{t\psi}\dot{t}+g_{\psi\psi}\dot{\psi}}{\eta(g_{rr}\dot{r}+g_{r\theta}\dot{\theta})
+\varepsilon(g_{r\theta}\dot{r}+g_{\theta\theta}\dot{\theta})}, \nonumber\\
\beta&=&r_{obs}\frac{p^{\hat{\theta}}}{p^{\hat{r}}}=
r_{obs}\frac{1}{\sqrt{g_{\theta\theta}}}\frac{g_{r\theta}\dot{r}+g_{\theta\theta}\dot{\theta}}{\eta(g_{rr}\dot{r}+g_{r\theta}\dot{\theta})
+\varepsilon(g_{r\theta}\dot{r}+g_{\theta\theta}\dot{\theta})},
\end{eqnarray}
where $r_{obs}, \theta_{obs}$ are the radial coordinate and polar angle of observer.
\begin{figure}[t]
\begin{center}
\includegraphics[width=4cm]{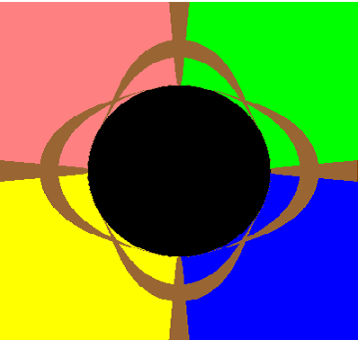}\includegraphics[width=4cm]{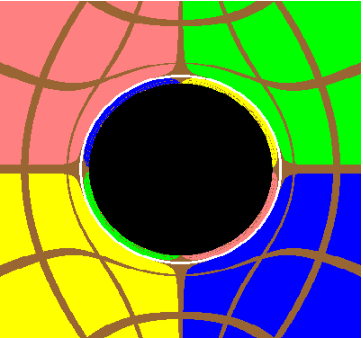}\includegraphics[width=4cm]{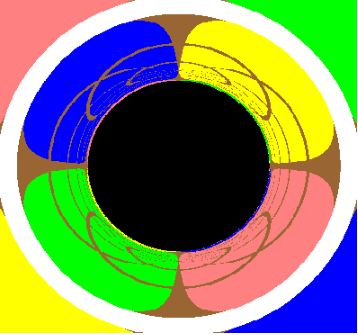}\includegraphics[width=4cm]{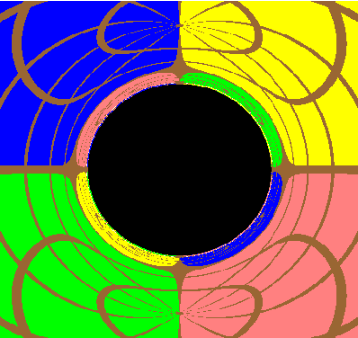}
\caption{ The variation of images with the deformation parameter $D_0$ for the circular disformal Kerr black hole with fixed $a=0$ and $X_0=-1$. Here we set the mass parameter $M=1$, $r_{obs}=30M$ and $\theta_{obs}=\pi/2$. The figures from left to right correspond to $D_0=-0.49$, $-0.3$, $0$, and $0.5$, respectively.}\label{Figa0thetapi/2}
\end{center}
\end{figure}

\begin{figure}[t]
\begin{center}
\includegraphics[width=4cm]{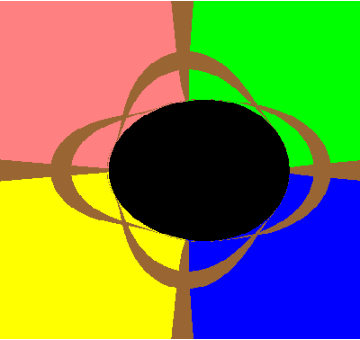}\includegraphics[width=4cm]{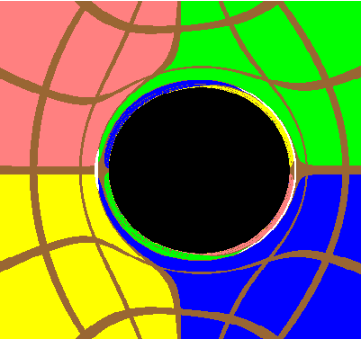}\includegraphics[width=4cm]{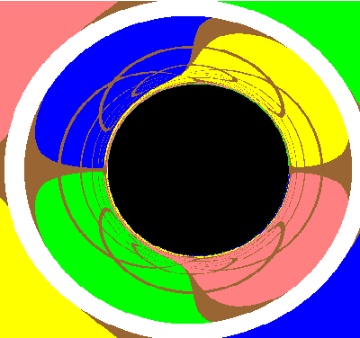}\includegraphics[width=4cm]{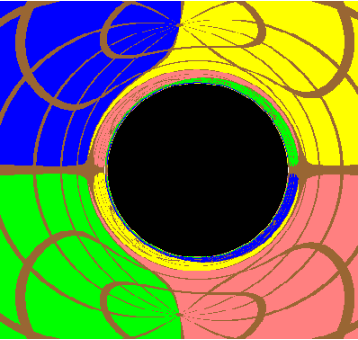}
\caption{The variation of images with the deformation parameter $D_0$ for the circular disformal Kerr black hole with fixed $a=0$ and $X_0=-1$. Here we set the mass parameter $M=1$, $r_{obs}=30M$ and $\theta_{obs}=\pi/2$. The figures from left to right correspond to $D_0=-0.49$, $-0.3$, $0$, and $0.5$, respectively.}\label{Figa05thetapi/2}
\end{center}
\end{figure}

\begin{figure}[t]
\begin{center}
\includegraphics[width=4cm]{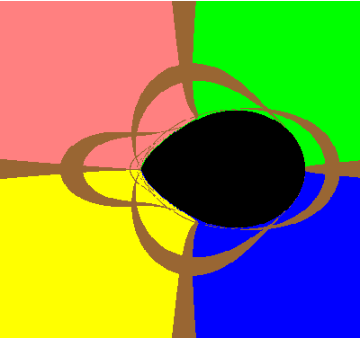}\includegraphics[width=4cm]{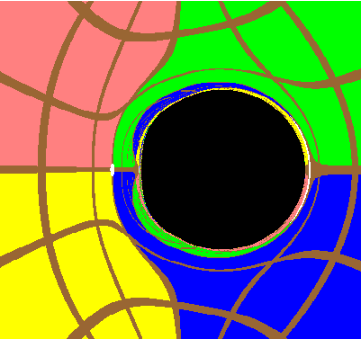}\includegraphics[width=4cm]{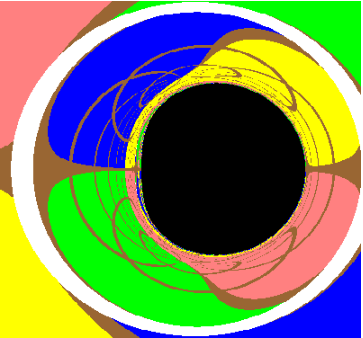}\includegraphics[width=4cm]{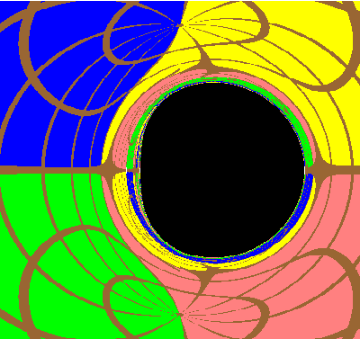}
\caption{ The variation of images with the deformation parameter $D_0$ for the circular disformal Kerr black hole with fixed $a=0$ and $X_0=-1$. Here we set the mass parameter $M=1$, $r_{obs}=30M$ and $\theta_{obs}=\pi/2$. The figures from left to right correspond to $D_0=-0.49$, $-0.3$, $0$, and $0.5$, respectively.}\label{Figa099thetapi/2}
\end{center}
\end{figure}

Figs.\ref{Figa0thetapi/2}-\ref{Figa099thetapi/2} present the images of the circular disformal Kerr black hole obtained by an observer located on the equatorial plane for different values of the rotation parameter $a$ and the deformation parameter $D_0$. To preserve the space-like nature of the radial direction outside the event horizon, one must impose the condition that $g_{rr}g_{\theta\theta}-g_{r\theta}^2>0$, which means $C_0>2D_0X_0$.
For the rotation parameter $a=0$, we find that the shadow remains a perfect circle with a constant radius $R=3\sqrt{3}$ for all values of $D_0$. This is because the black hole is static in this limit, and the deformation parameter appears only in the $g_{rr}$ component without changing the radius of the photon sphere. Consequently, the shadow size is independent of the deformation parameter $D_0$ and coincides with that in the Schwarzschild spacetime. The Einstein ring always remains smooth and circular. Its radius increases with the deformation parameter when $D_0>0$, while for $D_0<0$ the radius decreases as $D_0$ becomes smaller and eventually disappears.
For the rotating black hole, the size of the shadow decreases and its shape gradually becomes more flattened as the deformation parameter $D_0$ increases, while the shadow still maintains north-south symmetry. The change of the shadow shape becomes more distinct for the black hole with the more quickly rotation and the more negative parameter $D_0$, a behavior that is very similar to what occurs in the disformal Kerr black hole in quadratic DHOST theories\cite{W22}. Especially, as $a=0.99$ and $D_0<0$, the shadow eventually exhibit a very special ``almond" shape.

\begin{figure}[t]
\begin{center}
\includegraphics[width=3cm]{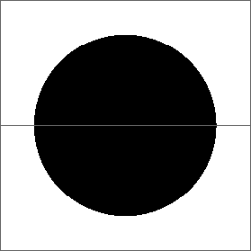}\includegraphics[width=3cm]{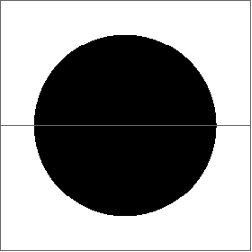}\includegraphics[width=3cm]{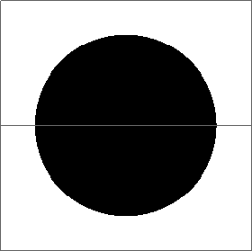}\includegraphics[width=3cm]{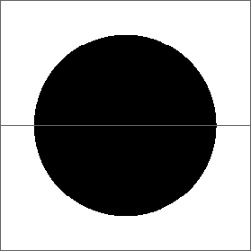}\includegraphics[width=3cm]{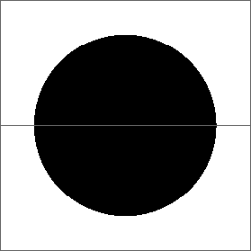}\\
\includegraphics[width=3cm]{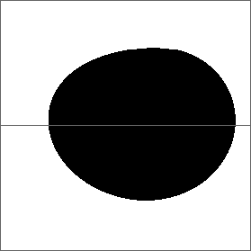}\includegraphics[width=3cm]{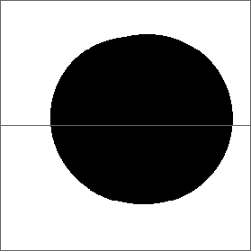}\includegraphics[width=3cm]{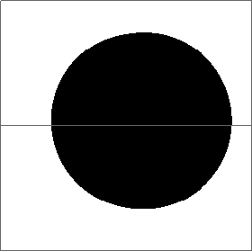}\includegraphics[width=3cm]{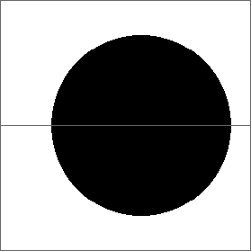}\includegraphics[width=3cm]{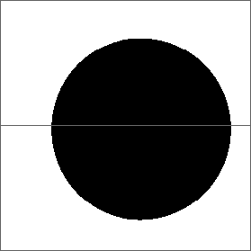}\\
\includegraphics[width=3cm]{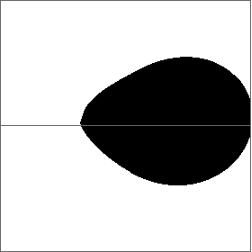}\includegraphics[width=3cm]{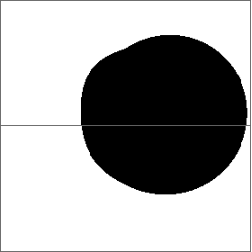}\includegraphics[width=3cm]{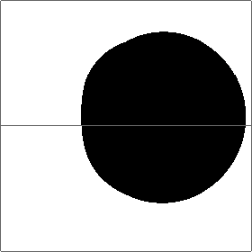}\includegraphics[width=3cm]{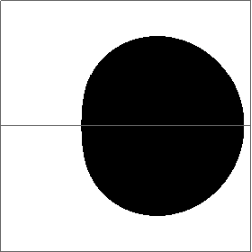}\includegraphics[width=3cm]{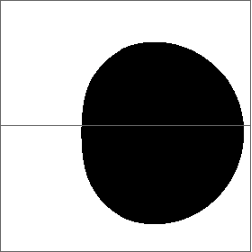}
\caption{ The variation of shadows with the disformal parameter $D_0$ for the circular disformal Kerr black hole with fixed $X_0=-1$. Here we set the mass parameter $M=1$, $r_{obs}=30M$ and $\theta_{obs}=\pi/3$. From top to bottom, the panels correspond to $a=0$, $a=0.5$ and $a=0.99$. From left to right, they correspond to $D0=-0.49$, $-0.4$, $-0.3$, $0$, and $0.5$.}\label{Figsmall}
\end{center}
\end{figure}

Figs.\ref{Figsmall} present the shadows of the circular disformal Kerr black hole obtained by an observer deviation from the equatorial plane with $\theta_{obs}=\pi/3$ for different values of the rotation parameter $a$ and the deformation parameter $D_0$. We find that effect of the deformation parameter $D_0$ on the shadow size is similar to that for an observer on the equatorial plane, i.e., the shadow remains a perfect circle with a constant radius for the non-rotating black hole, while in the rotating case its size becomes an increasing function of $D_0$. Compared with the case where the observer is located on the equatorial plane, the north-south symmetry of the rotating black hole shadow is broken when the observer moves away from the equatorial plane, due to the combined effects of the spin parameter and the $g_{r\theta}$ component. When the observer is located in the northern hemisphere, the shadow exhibits the following behavior. For $D_0<0$, as $D_0$ decreases further, the northern part of the shadow first expands and then contracts, while the southern part keeps contracting, and meanwhile the geometric center of the shadow shifts northward then southward. In contrast, for $D_0>0$, as $D_0$ increases, the northern part gradually contracts whereas the southern part expands, accompanied by a southward shift of the geometric center of the shadow. As a result, a pronounced asymmetry in the shadow shape emerges. When the observer is located in the southern hemisphere, these features are exactly reversed. These unique properties indicates that the circular disformal Kerr black hole can be distinguished from the Kerr and the disformal Kerr black hole in quadratic DHOST theories based on its image.

\begin{figure}[t]
\begin{center}
{\includegraphics[width=5.2cm]{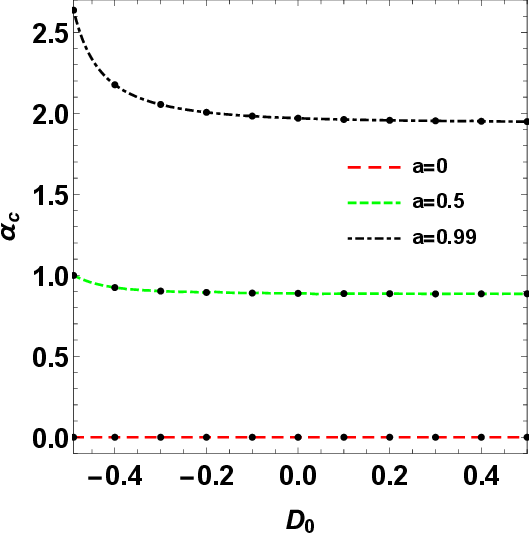}}\;\;{\includegraphics[width=5.3cm]{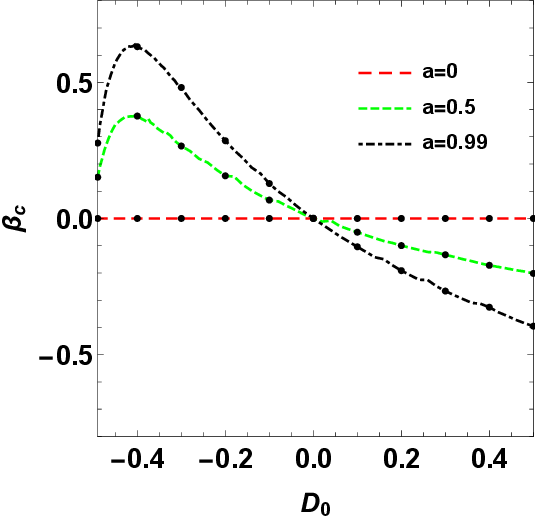}}\;\;{\includegraphics[width=5.2cm]{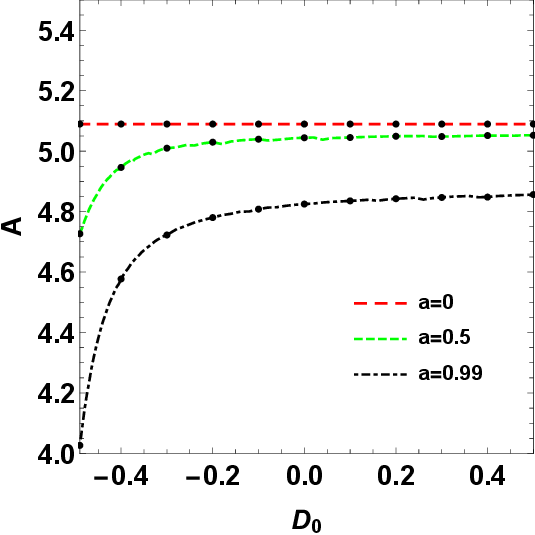}}
\caption{The change of the geometric center $(\alpha_c, \beta_c)$ and shadow area $A$ with the deformation parameter $D_0$ in the circular disformal Kerr black hole for different rotation parameter $a$. Here we set the mass parameter $M=1$, $r_{obs}=30M$, $\theta_{obs}=\pi/3$ and $X_0=-1$.}\label{XYA}
\end{center}
\end{figure}
In Fig.\ref{XYA}, we plot the variation of the geometric center $(\alpha_{c},\beta_{c})$ and the area $A$ of the shadow with respect to the deformation parameter $D_0$ in the circular disformal Kerr black hole for different rotation parameters $a$, for an observer located at $\theta_{obs}=\pi/3$. Here, the parameters $(\alpha_{c},\beta_{c})$ are defined as
\begin{equation}
\left(\alpha_c, \beta_c\right)=\frac{1}{N} \sum\left(\alpha_i, \beta_i\right),
\end{equation}
where $\left(\alpha_i, \beta_i\right)$ corresponds to the position of each point in the shadow region of the celestial coordinates and $N$ represents the number of pixels.
For the non-rotating case, the geometric center $(\alpha_{c},\beta_{c})$ of the shadow does not change with $D_0$ and the shadow remains a circle centered at the origin. For the rotating case,
For the deformation parameter $D_0<0$, a further decrease in $D_0$ causes the geometric center of the shadow to first shift toward the northeast and subsequently toward the southeast. In contrast, for $D_0>0$, an increase in $D_0$ leads to a southwestward shift of the geometric center, which is consistent with the previous analysis.
The area $A$ of the rotating black hole shadow increases as the deformation parameter $D_0$ increases, indicating that the size of the shadow becomes larger. Moreover, for the same value of $D_0$, the shadow area in the rotating case is always smaller than that in the non-rotating case.
From Fig.\ref{XYA}, we confirm again that the change of the shadow center with $D_0$ becomes more distinct for the rapidly rotating black hole.
These new shadow features arising from the scalar field help us understand the circular disformal Kerr black hole and provide a theoretical basis for testing the nature of gravity in the future.

\section{Summary}
In this paper we have investigated the shadows of the circular disformal Kerr black hole with the deformation parameter $D_0$, analyzed the influence of parameter $D_0$ on the shape and the geometric center of the shadow. Our result show that for the rotation parameter $a=0$, the shape of the shadow remains a perfect circle and coincides with that in the Schwarzschild spacetime for all values of the deformation parameter $D_0$. For the rotating case, the size of the shadow decreases as the deformation parameter decreases, and the shadow gradually becomes more flattened. It is worth noting that when the observer lies on the equatorial plane, the observed black hole shadow of the rotating black hole always exhibits north-south symmetry. However, once the observer deviation from the equatorial plane, this symmetry is broken. When the observer is located in the northern hemisphere, for $D_0<0$, a further decrease in $D_0$ causes the northern edge of the shadow to first expand and then contract, while the southern edge continues to contract, resulting in the geometric center first shifting northward and then reversing to a southward shift. In contrast, for $D_0>0$, increasing $D_0$ results in a contraction of the northern edge and an expansion of the southern edge, with the geometric center shifting southward. The situation is reversed for observers in the southern hemisphere. We further verify that the influence of $D_0$ on the position of the shadow center is significantly enhanced for rapidly rotating black holes. Especially, as $a=0.99$ and $D_0<0$, the shadow eventually exhibit a very special ``almond" shape. These results help clarify how disformal modifications influence black hole shadow formation, thereby providing useful theoretical guidance for probing possible deviations from general relativity with future high-resolution astronomical observations.

\section{\bf Acknowledgments}
This work was partially supported by the National Natural
Science Foundation of China (Grant Nos. 12205140, 12275078, 11875026, 12035005, and 2020YFC2201400)
and the Natural Science Foundation of Hunan Province (Grant No. 2023JJ40523).

\end{document}